\documentclass[12pt]{article}

\usepackage{amssymb,amsmath,amsfonts,eurosym,geometry,ulem,graphicx,color,setspace,sectsty,comment,footmisc,caption,pdflscape,subfigure,array,hyperref}
\usepackage[toc,page]{appendix}
\usepackage[unnumberedbib,bibnewpage]{apacite}
\usepackage{xcolor}
\usepackage{listings}
\usepackage{setspace}
\usepackage{bbm}
\usepackage[ font=footnotesize, justification = centering]{caption}

\newcolumntype{L}[1]{>{\raggedright\let\newline\\arraybackslash\hspace{0pt}}m{#1}}
\newcolumntype{C}[1]{>{\centering\let\newline\\arraybackslash\hspace{0pt}}m{#1}}
\newcolumntype{R}[1]{>{\raggedleft\let\newline\\arraybackslash\hspace{0pt}}m{#1}}

\def\code#1{\texttt{#1}}

\DeclareMathOperator*{\argmax}{arg\,max}

\begin{document}

\begin{titlepage}
\title{Calibrated Projection in MATLAB: Users' Manual\thanks{We gratefully acknowledge financial support through NSF grants SES-1230071 (Kaido), SES-0922330 (Molinari), and SES-1260980 (Stoye).}}
\author{\normalsize{Hiroaki Kaido}\thanks{Department of Economics, Boston University, hkaido@bu.edu.} \and
            \normalsize{Francesca Molinari}\thanks{Department of Economics, Cornell University, fm72@cornell.edu.} \and
            \normalsize{J\"org Stoye}\thanks{ Departments of Economics, Cornell University and University of Bonn, stoye@cornell.edu.} \and
            \normalsize{Matthew Thirkettle}\thanks{Department of Economics, Cornell University, mkt68@cornell.edu.}}
\date{\normalsize{\today}}
\maketitle
\begin{abstract}
\noindent We present the calibrated-projection MATLAB package implementing the method to construct confidence intervals proposed by \citeA{kaido2017confidence}.  This manual provides details on how to use the package for inference on projections of partially identified parameters.  It also explains how to use the MATLAB functions we developed to compute confidence intervals on solutions of nonlinear optimization problems with estimated constraints. \\
\vspace{0in}
\bigskip
\end{abstract}
\setcounter{page}{0}
\thispagestyle{empty}
\end{titlepage}
\pagebreak \newpage

\doublespacing

\section{Introduction} \label{sec:introduction}
This manual details the structure of the Calibrated Projection Interval (CPI) algorithm and \code{MATLAB} Package.  It accompanies the paper ``Confidence Intervals for Projections of Partially Identified Parameters'' \cite{kaido2017confidence} and it assumes familiarity with that paper.\footnote{Some notation differs between this paper and \cite{kaido2017confidence}.  This is made clear throughout this manual.  Unless otherwise specified, we use notation from the earlier version of the paper \cite{kaido2016confidence}.  The table numbering references  \cite{kaido2017confidence}.}  The CPI algorithm uses an EAM (evaluate, approximate, maximize) algorithm to solve:
\begin{align*}
& \inf/\sup_{\theta \in \Theta} p'\theta \\
& \text{s.t.} \sqrt{n} \frac{\bar m_{j}(\theta)}{\hat \sigma_{j}(\theta)} \leq \hat c(\theta) \quad j=1,\cdots,J,
\end{align*}
where $\hat c(\theta)$ is the calibrated critical value \cite{jones1998efficient,jones2001taxonomy}.  This version of the CPI algorithm is optimized for basis projection $p = (0,\cdots,0,1,0,\cdots,0)$ with hyperrectangle parameter constraints $\Theta = \{\theta \in \mathbb{R}^d: \theta_{LB} \leq \theta \leq  \theta_{UB}\}$.  We also allow for $p$ to be in the unit sphere and polytope constraints on the parameter space, so that
$\Theta = \{\theta \in \mathbb{R}^d: \theta_{LB} \leq \theta \leq  \theta_{UB}, A_{\theta} \theta \leq b_{\theta}\}$.  Additional care is required within these extensions (see Appendix \ref{appendix:polytope} for further details).   The current version of the package is written for moment (in)equalities that are separable in data $W$ and parameter $\theta$, so that $E_P[m_j(W_i,\theta)] = E_P[f_j(W_i)] + g_j(\theta)$.\footnote{In this manual and in the CPI MATLAB package data is defined as $W$.  The function $f$ and $g$ are the two components of the separable moment (in)equality $E_P[m_j(W_i,\theta)]$.  This is in contrast to \citeA{kaido2017confidence}, where data is $X$, $f(\theta)$ refers to the objective function, and $\bar g(\theta)$ appears in the EAM algorithm.  The subscript $n$ has also been dropped from all estimators.} Future releases of the package will include:
\begin{itemize}
\item Non-separability of $E_P[m_j(W_i,\theta)]$ in $W_i$ and $\theta$.
\item Objective function $h(\theta)$ not necessarily equal to $p'\theta$.
\end{itemize}
We have structured the code so that it is portable.  In order to implement a user-specified model, the user needs only input the data, algorithm options, the function that defines the estimators for the moment (in)equalities, as well as the gradients and standard deviation estimators of the moment functions.    Section \ref{sec:using} details how to use the portable code with a user-specified partially identified model with separable moment inequalities.  Section \ref{sec:using} also provides instructions on how to replicate the simulations in \citeA{kaido2017confidence}.     Section \ref{sec:algorithm} provides a deeper insight into how the CPI algorithm is structured. Section \ref{sec:conclusion} discusses extensions to the algorithm and concludes.

\section{Using the Calibrated Projection Interval Algorithm} \label{sec:using}
In this section we detail the steps required to run the simulations in \citeA{kaido2017confidence}, and how to run  a user-specified model.  We use as a working example the Entry Game Model in Section 5 of \citeA{kaido2017confidence}.   This section is organized as follows.  Section \ref{sec:overview} briefly describes the key files in the package.  Section \ref{subsec:cvxgen} details how to set up \code{CVXGEN} and \code{CVX}, both are fast disciplined convex solvers that we use to compute the calibrated critical value $\hat c(\theta)$ \cite{mattingley2012cvxgen,cvx,gb08}.  Section \ref{sec:simulations} provides instructions on how to replicate the simulations to reproduce the tables in \citeA{kaido2017confidence}.  Section \ref{sec:usermodel} provides instructions on how to implement a user-specified model and compute Projection Intervals (either Calibrated or Andrew and Soares (AS) \cite{andrews2010inference}) using the CPI algorithm.

\subsection{Overview of Important Files and Folders}\label{sec:overview}
First, we briefly describe the key \code{MATLAB} files and folders.
\begin{itemize}
\item \code{KMS\_Simulation.m}.  This executes the simulations in  \citeA{kaido2017confidence}.  The DGP, method (Calibrated Projection, Andrew and Soares (AS), or Bugni, Canay, and Shi (BCS)-Profiling),\footnote{The code implementing BCS is the code provided by these authors and is available at \url{http://qeconomics.org/ojs/index.php/qe/article/view/431}.} nominal significance level, projection directional vector, number of observations, and number of simulations are set by the user here.  The data is generated and passed to either \code{KMS\_0\_Main.m} or \code{BCS\_Main}, which computes the Calibrated or AS Projection Interval, or the BCS-Profiled Interval, respectively.

\item \code{KMS\_0\_Main.m}.  This is the file that the user calls to execute the CPI algorithm and compute the Projection Interval (either Calibrated or AS).  The user specifies data \code{W}, the initial guess for a feasible parameter \code{theta\_0}, the projection direction \code{p}, a set of pre-specified feasible points \code{theta\_feas}, the lower bound on parameter space \code{LB\_theta}, the upper bound on parameter space \code{UB\_theta},  the polytope constraints on the parameter space \code{A\_theta} and \code{b\_theta} so that $A_{\theta} \theta \leq b_{\theta}$, the nominal significance level \code{alpha}, a one-sided or two-sided confidence interval \code{type}, the projection method (calibrated or AS) \code{CI\_method}, the GMS tuning parameter \code{kappa}, the GMS function \code{phi}, the name of the \code{MEX} files for \code{CVXGEN} (discussed in Section \ref{subsec:cvxgen} below) \code{CVXGEN\_name}, and a structure of algorithm options \code{KMSoptions}.
\end{itemize}
 The package assumes that the moment (in)equalities are separable, so that $E_P[m_j(W_i,\theta)] = E_P[f_j(W_i)] + g_j(\theta)$.
\begin{itemize}
\item \code{moments\_w.m} is the user-specified function for the estimator of $E_P[f_j(W_i)]$, namely $\hat f_{j}$. We allow for both moment inequalities and equalities, as well as paired moment inequalities.  If $f_j(W_i)$ is a Bernoulli random variable and if its expectation is too close to $0$ or $1$, then the corresponding moment (in)equalities are dropped.  The output \code{f\_ineq\_keep} and \code{f\_eq\_keep} defines the moment (in)equalities that are not discarded.

\item \code{moments\_theta.m} is the user-specified function for $g_j(\theta)$.

\item \code{moments\_gradient.m} is the user-specified function for the gradient of $g_j(\theta)$, which is denoted $D_{\theta}g_j(\theta)$.

\item \code{moments\_stdev.m} is the user-specified function for the estimator for the standard deviation $\sigma_j(W_i)$.

\item \code{KMSoptions.m} defines a structure of algorithm options.  \code{KMSoptions} is also passed to the four user-specified functions above, so the user can pass additional parameters through \code{KMSoptions} to the user-specified functions (e.g., the support for data $W_i$).  The function \code{KMSoptions.m} is called before running \code{KMS\_0\_Main.m}, and is passed through the last  argument of \code{KMS\_0\_Main.m}, which is \code{KMSoptions}.

\item \code{Rho\_Polytope\_Box.m} and \code{bound\_transform.m} are additional user-written functions needed when polytope constraints on the parameter space are provided (see  the arguments \code{A\_theta} and \code{b\_theta} in \code{KMS\_0\_Main.m}) or when $p$ is not a basis vector.  If $p$ is a non-basis vector or if polytope constraints on the parameter space are included, then sensitivity in the estimate for the projection interval can arise.  Details about these files, as well as more detail on the sensitivity issues, is discussed in Appendix \ref{appendix:polytope}.
\end{itemize}
The disciplined convex solver \code{CVXGEN}  is used to check whether the set
\[
\Lambda^b(\theta,\rho,c) = \{ \lambda \in \sqrt{n}(\Theta-\theta) \cap \rho B^d : \mathbb{G}^b_{j} +  D_{\theta}g_j(\theta)\lambda + \varphi_j(\hat \xi_{j}(\theta)) \leq c, j = 1,\cdots,J\}
\]
is empty for each bootstrap repetition $b = 1,\cdots,B$.  In order to run \code{CVXGEN}, the user first compiles a  \code{MEX} file that defines the parameters of the problem (details in Section \ref{subsec:cvxgen}).
\begin{itemize}
\item The compiled \code{MEX} files are stored in the subfolder \code{\textbackslash CVXGEN}.  The file name for this is chosen by the user.  For example, we choose \code{csolve\_DGP8.mex64} for the BCS Entry Game.  The file name must also be defined when  \code{KMS\_0\_Main.m} is called.  The name is passed via the argument \code{CVXGEN\_name}.
\end{itemize}

\subsection{CVXGEN and CVX Setup}\label{subsec:cvxgen}
The calibrated critical value $\hat c(\theta)$ is computed using a fixed-point algorithm.  The fixed-point mapping is computed by checking whether the following set is empty:
\begin{align}\label{eq:lambda}
\Lambda^b(\theta,\rho,c) = \{ \lambda \in \sqrt{n}(\Theta-\theta) \cap \rho B^d : \mathbb{G}^b_{j}(\theta) +  D_{\theta}g_j(\theta)\lambda + \varphi_j(\hat \xi_{j}(\theta)) \leq c, j = 1,\cdots,J\}.
\end{align}
This amounts to solving many linear programs (LP), which is done using the fast disciplined convex solver \code{CVXGEN} \cite{mattingley2012cvxgen} or \code{CVX} \cite{cvx,gb08}.

\subsubsection{CVXGEN Setup}
To set up \code{CVXGEN}, the user needs to: 1) install a \code{MEX} Compiler; 2) generate \code{C} code at  \url{https://cvxgen.com};   3) compile and save the \code{MEX} file; 4) Instruct the CPI algorithm to use \code{CVXGEN} rather than \code{CVX}.

The first step is to install a \code{MEX} compiler.  We use the \code{MinGW-w64 Compiler} on a Windows machine, which is an add-on in \code{MATLAB}.  To install: open \code{MATLAB}, go to \code{Home} tab, go to \code{Add-Ons}.  An add-on search window appears on the screen.  Search \code{MinGW-w64 Compiler} and install \code{MATLAB Support for MinGW-w64 C/C++ Compiler v}. On a Mac, a \code{C} compiler is supplied with \code{Xcode}.  On a Linux based system, one can use \code{GCC} (GNU Compiler Collection).

The next step is to generate the \code{C} code for a specific problem.  First, create an account at \url{https://cvxgen.com} and log in.  Next, navigate to the \code{edit} tab under \code{problem}.  Copy-and-paste the following:

\footnotesize
\begin{lstlisting}[backgroundcolor = \color{gray!30},
                   xleftmargin = 0cm,
                   framexleftmargin = 1em]
dimensions
    dim_p     = XX
    J1        = YY
    J2        = ZZ
    S         = VV
end
parameters
    A   (J1 + 2*J2 + 2*dim_p + 2 + S , dim_p)
    b   (J1 + 2*J2 + 2*dim_p + 2 + S , 1)
end
variables
    x(dim_p,1)
end
minimize
    0
subject to
    A*x<= b
end
\end{lstlisting}

 \normalsize
​Replace \code{XX} with the dimension of the parameter $\theta$, \code{YY} with the number of moment inequalities, \code{ZZ} with the number of moment equalities (do not double count $E_P[m_j(W_i,\theta)] \leq 0$ and $-E_P[m_j(W_i,\theta)] \leq 0$ here), and \code{VV} with the number of polytope box constraints.  If no polytope constraints $A_{\theta} \theta \leq b_{\theta}$ are included, set \code{VV}$=0$.

Next, navigate to the \code{generate C} tab under \code{CODEGEN}.  Click \code{Generate code}.  As a result, a list of files populate the webpage.  Download the \code{cvxgen.zip} file and extract.  Run \code{make\_csolve.m}.  The file \code{csolve.mex64} should appear in the folder (if on a Linux or Mac machine, the extension is slightly different).\footnote{If an error occurs here, it is likely that the \code{MEX} compiler is not installed correctly.}  Rename \code{csolve.mex64} to \code{CVXGEN\_name.mex64} (where \code{CVXGEN\_name} is specified by the user) and move the file to the subfolder  \code{\textbackslash CVXGEN}.

Last, set \code{KMSoptions.CVXGEN = 1} to instruct CPI algorithm to use \code{CVXGEN}.

There is an upper bound of $4,000$ non-zero Karush-Kuhn-Tucker matrix entries for the linear program in \code{CVXGEN}.  The size of the problem is determined jointly by $J_1$, $J_2$, and $d$.  As an example, \code{CVXGEN} can handle $\theta \in \mathbb{R}^{10}$ with $J_1 = 55$ and $J_2 = 55$.

\subsubsection{CVX Setup}
An alternative solver to \code{CVXGEN} is \code{CVX}.  This solver is slower than \code{CVXGEN}, but can handle significantly larger LPs and, in our experience, is significantly faster than \code{MATLAB}'s LP solver \code{LINPROG}.  \code{CVX} is a \code{MATLAB} ``wrapper'' for five different disciplined convex solvers \cite{cvx,gb08}.  Among these, the solver \code{MOSEK} is the fastest for our problem.  To run \code{CVX} with \code{MOSEK}:
\begin{enumerate}
\item Ensure that there is a copy of \code{CVX} is located in the subfolder \code{\textbackslash CVX}.  If not, navigate to \url{http://cvxr.com/cvx/} and deposit a copy in the subfolder \code{\textbackslash CVX}.
\item Request a license from  \url{http://cvxr.com/cvx/} and deposit it in the same folder.
\item Run \code{cvx\_setup.m}.
\item Set solver using the command \code{cvx\_solver MOSEK} in the \code{MATLAB} command window.
\item Set \code{KMSoptions.CVXGEN = 0}.
\item Set  \code{CVXGEN\_name} to the empty set.
\end{enumerate}
Once \code{CVXGEN} or \code{CVX} is set up, either a simulation model (Section \ref{sec:simulations}) or a user-specified model (Section \ref{sec:usermodel}) can be called via the CPI algorithm.\footnote{For additional help with \code{CVXGEN} or \code{CVX}, please visit   \url{https://cvxgen.com} and \url{http://cvxr.com/cvx/}.}

\subsection{Running Simulations}\label{sec:simulations}
In this section we discuss how to replicate the simulation results in \citeA{kaido2017confidence} (see Tables 1-7 in the paper).  As per \code{CVXGEN} policy, we are unable to distribute the \code{MEX} files for these simulations.  So the first step is to generate the relevant \code{MEX} files, see Section \ref{subsec:cvxgen} for instructions and Table \ref{tab:cvxgenparams} for \code{CVXGEN} parameters and naming conventions.

The next step is to set parameters in \code{KMS\_Simulation.m}.  Open an instance of \code{KMS\_Simul\-ation.m} and set the following:
\begin{itemize}
\item \code{method = 'KMS'} to compute the Calibrated Projection Interval; or \code{method = 'AS'} to compute the AS Projection Interval.
\item \code{DGP=k} where \code{k}$\in \{1,2,3,4,5,6,7,8\}$. This parameter selects the data-generating process.  $k=1-4$ corresponds to the rotated box described in the earlier version \citeA{kaido2016confidence}. $k=5-8$ corresponds to the Entry Games: $k=5$ is the point-identified Entry Game with zero correlation (Table 3); $k=6$ is the partially-identified Entry Game with zero correlation (Tables 4, 6 and 7); $k=7$ is the partially-identified Entry Game with $Corr(u_1,u_2) = 0.5$ (Table 5);     $k=8$ is the BCS simulation (Tables 1-2 in \citeA{kaido2017confidence}).
\item \code{KMS=1} or \code{KMS=0} determines if \code{KMS\_0\_Main} or \code{BCS\_Main} is called. \code{KMS=0} is a valid input only if \code{DGP=8}, and  \code{component=1} or \code{component=2}.
\item \code{component=k} where \code{k}$\in \{1,\cdots,\text{\code{dim\_p}}\}$ selects the projection direction.  That is, the projection vector is $p$ with $p_i = 1$ if $i=k$ and $p_i = 0$ otherwise.
\item \code{n} is the sample size.  \code{n} is set to $4000$ for Tables 1-7.
\item \code{Nmc} is the number of Monte Carlo simulations requested.  \code{Nmc} is set to 300 in Table 1 and 1000 in Tables 2-7.
\item \code{sim\_lo} and \code{sim\_hi} determine which simulations are run.  These parameters are used to split the simulations into batches if needed.
\end{itemize}
Among other things, convergence criteria  are set in \code{KMSoptions}.  All DGPs other than the correlated error DGP, which is DGP 7, use what we call the baseline options.  DGP 7, on the other hand, is a fairly difficult problem to solve.  Therefore, we use more stringent convergence criteria for this DGP.  The options listed below and, in particular, the contraction rates are discussed in more detail in Section \ref{sec:EAM}.

The baseline options are:
\begin{itemize}
\item  \code{KMSoptions.EAM\_maxit=20}. This sets the maximum number of EAM iterations to 20.
\item \code{KMSoptions.h\_rate=1.8}. This determines the contraction rate of the parameter space for the M-step.
\item \code{KMSoptions.h\_rate2=1.25}.  This determines the contraction rate of the parameter space for additional points
\item \code{KMSoptions.EAM\_obj\_tol  = 0.005}.  One requirement for convergence is that the absolute difference between the expected improvement projection and the current feasible point $\theta^{*,L}$ is less than  \code{EAM\_obj\_tol}.
\item \code{KMSoptions.EI\_points=10}.  The M step is initialized with a set of starting points.  The algorithm selects \code{EI\_points} points around the current feasible point $\theta^{*,L}$ that have positive expected improvement.  Additional points are also selected.
\end{itemize}
The stringent options for \code{DGP=7} are:
\begin{itemize}
\item  \code{KMSoptions.EAM\_maxit=50}.
\item \code{KMSoptions.h\_rate=1.25}.
\item \code{KMSoptions.h\_rate2=1.15}.
\item \code{KMSoptions.EAM\_obj\_tol  = 0.0001}.
\item \code{KMSoptions.EI\_points=20}.
\end{itemize}
The number of bootstrap repetitions is also set in \code{KMSoptions.m}.  Table 1 sets this number equal to 301, so that \code{KMSoptions.B=301}.  For Tables 2-7 set \code{KMSoptions.B=1001}.

Finally, run \code{KMS\_Simulation} to run a simulation with the parameters and options specified above.  The results are saved in the subfolder \code{\textbackslash Results}.

The file \code{Analysis.m} carries out post analysis for a particular set of simulations.  To run the post analysis, load a results file and run \code{Analysis.m}.  The output includes the median lower bound for the Calibrated Projection Interval; the median upper bound for the Calibrated Projection Interval; coverage percent at the end points of the identification region, as well as at the true parameter; average $\hat c(\theta)$; and average computational time.

\subsection{User-specified Model}\label{sec:usermodel}
In this section we detail the files that need to be modified so that a user can implement the algorithm to compute the Calibrated Projection Interval for a user-specified problem.  We use \code{DGP=6} as a working example (this is the partially-identified Entry Game with zero correlation between the unobservable shocks $u_1$ and $u_2$).  For tractability, we have created a working example file \code{ExampleDGP6.m}.  Here, we generate one data set with $n=4000$ observations.  We assume that the parameter space is a hyperrectangle and $p$ is a basis vector throughout the rest of this section.

\subsubsection*{Step 1: Set up \code{CVXGEN} or \code{CVX}}

The first step is to set up either \code{CVXGEN} or \code{CVX}.  This is described in detail in Section \ref{subsec:cvxgen}.  The required parameters are the dimension of the parameter $\theta$, \code{dim\_p}, the number of moment inequalities, \code{J1}, and the number of moment equalities, \code{J2}.  In our working example, set \code{dim\_p} = 8, \code{J1 = 8}, and \code{J2 = 8}.  Name the \code{MEX} file as \code{exampleDGP6.mex64} and deposit it in the subfolder \code{\textbackslash CVXGEN}.  Set \code{KMSoptions.CVXGEN = 1}.  Alternatively, if \code{CVX} is used, follow the installation instructions at the end of Section \ref{subsec:cvxgen} and set \code{KMSoptions.CVXGEN = 0}.

\subsubsection*{Step 2: Moment (In)equality: Estimator for $E_P[f_j(W_i)]$}

The moment (in)equalities are separable so that $E_P[m_j(W_i,\theta)] = E_P[f_j(W_i)] + g_j(\theta)$ for $j=1,\cdots,J$.  The estimator for $E_P[f_j(W_i)]$ is defined in the function \code{moments\_w.m}.  Override the  file \code{moments\_w.m} with the following shell:

\footnotesize
\begin{lstlisting}[backgroundcolor = \color{gray!30},
                   xleftmargin = 0cm,
                   framexleftmargin = 1em]
function [f_ineq,f_eq,f_ineq_keep,f_eq_keep,paired_mom,J1,J2,J3] ...
            = moments_w(W,KMSoptions)
    f_keep_threshold  = KMSoptions.f_keep_threshold;
    J1                = XX1;
    J2                = XX2;
    J3                = XX3;
    f_ineq            = zeros(J1,1);
    f_eq              = zeros(2*J2,1);
    f_ineq_keep       = zeros(J1,1);
    f_eq_keep         = zeros(2*J2,1);
    paired_mom        = zeros(J3,1);
%% Define output here %%
end
\end{lstlisting} \normalsize
The inputs are data \code{W}, which is $n$-by-$d_W$ and \code{KMSoptions}.  The outputs are:
\begin{enumerate}
\item \code{f\_ineq} is a $J_1$-by-$1$ vector of moment inequalities.  As an example, consider \code{Lines 208-212} in \code{moments\_w.m}, which correspond to the Entry Game moment inequalities (See Equations 5.3-5.4 in \citeA{kaido2017confidence}):
    \footnotesize
    \begin{align*}
    E_P[m_{j}(W_i,\theta)] & = E_P[ f_{j}(W_i)] + g_{j}(\theta), \quad \text{with} \\
    E_P[ f_{j}(W_i)] & = E_P[\mathbbm{1}(Y = (0,1)) \mathbbm{1}(X = x)] \quad \text{and} \\
    g_{j}(\theta) &=  - G_r((-\infty,-x_1'(\beta_1 + \Delta_1) \times [-x_2'\beta_2,\infty)))p_x\\
     E_P[m_{j+1}(W_i,\theta)] & =E_P[ f_{j+1}(W_i)]  + g_{j+1}(\theta), \quad \text{with}   \\
  E_P[ f_{j+1}(W_i)] & = -E_P[\mathbbm{1}(Y = (0,1)) \mathbbm{1}(X = x)] \quad \text{and} \\
    g_{j+1}(\theta) & = [G_r((-\infty,-x_1'(\beta_1 + \Delta_1) \times [-x_2'\beta_2,\infty))  \\
    & \quad \quad -G_r((-x_1'\beta_1,-x_1'(\beta_1 + \Delta_1) \times [-x_2'\beta_2,-x_2'(\beta_2 + \delta_2)))]p_x,
    \end{align*}\normalsize
    where $W = (Y,X)$.\footnote{We define data $W$ with firm-entry decisions $Y$ and market characteristics $X$. \citeA{kaido2017confidence} define data $X$ with firm-entry decisions $Y$ and market characteristics $Z$.} Moment inequalities $j \in \{1,3,5,7\}$ and $j+1$ correspond to a point $x$ in the support $\mathcal{X} \equiv \{(-1,-1),(-1,1),(1,-1),(-1,-1)\}$.  For each $x \in \mathcal{X}$,  $E_P[ f_{j+k}(W_i)]$, $k=0,1$,  is estimated by:
    \begin{align*}
  \hat  f_{j} & = \frac{1}{n} \sum_{i=1}^n \mathbbm{1}(Y_i = (0,1)) \mathbbm{1}(X_i = x) \\
   \hat f_{j+1} & = -\frac{1}{n} \sum_{i=1}^n \mathbbm{1}(Y_i = (0,1)) \mathbbm{1}(X_i = x).
    \end{align*}
    This corresponds to \code{Lines 208-212} in \code{moments\_w.m}:
    \footnotesize
    \begin{lstlisting}[backgroundcolor = \color{gray!30},
                   xleftmargin = 0cm,
                   framexleftmargin = 1em]
f_ineq((ii-1)*2 + 1,1) = sum(Y1 == 0 & Y2 == 1  & X1 == x1 & X2 == x2)/n;
f_ineq((ii-1)*2 + 2,1) = - sum(Y1 == 0 & Y2 == 1  & X1 == x1 & X2 == x2)/n;
    \end{lstlisting}\normalsize

    \item \code{f\_eq} is a $2 J_2$-by-$1$ vector of moment equalities.  Entries $j=1,\cdots,J_2$ of  \code{f\_eq} are defined on \code{Lines 215-218} in \code{moments\_w.m} and correspond to the moment equalities with $E_P[m_j(W,\theta)] \leq 0$.  Entries $j=J_2+1,\cdots,2J_2$ of  \code{f\_eq} correspond to  $-E_P[m_j(W,\theta)] \leq 0$.  It is important to include both the positive and negative of $E_P[m_j(X,\theta)]$ for the moment equalities, see \code{Line 222} in \code{moments\_w.m}: \code{ f\_eq = [f\_eq ; -f\_eq];}.

    \item  \code{f\_ineq\_keep} and  \code{f\_eq\_keep} are $J_1$-by-$1$ and $2 J_2$-by-$1$ vectors of indicators.  These define which moment (in)equalities we keep.  If $f_j(W)$ has unbounded support, then set the corresponding entry in $\code{f\_ineq\_keep}$ and $\code{f\_eq\_keep}$ equal to 1 (see the rotated box example on \code{Lines 87-88} in \code{moments\_w.m}).  In the Games Example, $f_j(W_i)$ is bounded by $0$ and $1$. If  $E_P[f_j(W_i)]$ is close to $0$ or $1$, Assumption 4.1-(iv) in \citeA{kaido2017confidence}, which is taken from \citeA{andrews2010inference} and common in the literature, is violated. Therefore, if $\hat f_{j}$ is within the tolerance of \code{KMSoptions.f\_keep\_threshold} of $0$ or $1$, then the corresponding component of  \code{f\_ineq\_keep} or \code{f\_eq\_keep} is set equal to 0, indicating that moment is dropped completely from the analysis -- it is not used to compute $\hat c(\theta)$ and it does not enter the M-step (see Section \ref{sec:EAM}, Pg 24).\footnote{In the Games Example, the moment inequality functions $f_j(W_i)$  for $j \in \{2,4,6,8\}$ are bounded by $-1$ and $0$.  Thus moment inequality $j  \in \{2,4,6,8\}$ is dropped for the analysis if $\hat f_j$ is too close to $-1$ or $0$.}   Otherwise it is set to 1.  See \code{Lines 235-236} in \code{moments\_w.m}.  \code{KMSoptions.f\_keep\_threshold} is a user-specified option with default value equal to $10^{-4}$ and may be modified in different applications.

    \item \code{paired\_mom} is a $J_3$-by-$1$ vector indicating the paired moment inequalities.  If there are no paired moment inequalities, set  \code{paired\_mom} to the empty set.  For each paired moment inequalities, set the corresponding elements in \code{paired\_mom} equal to a unique indicator $j = 1,\cdots,J_3$.  See \code{Lines 227-230}  in \code{moments\_w.m}.

    \item $J_1$, $J_2$, and $J_3$ define the number of moment inequalities, equalities, and paired moment inequalities.
\end{enumerate}
In the shell above, replace \code{XX1} - \code{XX3} with the number of moment inequalities, moment equalities, and paired moment inequalities.  Preset each output to the zero vector or the empty set as described above.  Last, input user-specified functions for each output.

\subsubsection*{Step 3: Moment (In)equality: Model-implied Moment Function $g_j(\theta)$}
The model-implied function $g_j(\theta)$ is defined in the function \code{moments\_theta.m}.  Override  \code{moments\_theta.m} with the following shell:
\footnotesize
\begin{lstlisting}[backgroundcolor = \color{gray!30},
                   xleftmargin = 0cm,
                   framexleftmargin = 1em]
function [g_ineq,g_eq] = moments_theta(theta,J1,J2,KMSoptions)
    g_ineq            = zeros(J1,1);
    g_eq              = zeros(2*J2,1);
%% Define output here %%
end
\end{lstlisting} \normalsize
The inputs are the $d$-by-$1$ parameter vector \code{theta}, number of moment inequalities \code{J1}, number of moment equalities \code{J2}, and algorithm options \code{KMSoptions}.  The outputs are the $J_1$-by-$1$ vector of moment inequalities \code{g\_ineq} and the $2J_2$-by-$1$ vector of moment equalities \code{g\_eq}, where entries $j=1,\cdots,J_2$ of \code{g\_eq} correspond to $E_P[m_j(W_i,\theta)] \leq 0$ and entries $j = J_2+1,\cdots,2J_2$ of \code{g\_eq} correspond to $-E_P[m_j(W_i,\theta)] \leq 0$.

In the shell above, input user-specified functions for outputs \code{g\_ineq} and \code{g\_eq}.

As an example, consider \code{DGP=6}.  The moment functions are defined on \code{Lines 100-169} in  \code{moments\_theta.m}.  For example, the moment inequality in Equation (5.3) in \citeA{kaido2017confidence} is:
    \footnotesize
    \begin{align*}
    E_P[m_{j}(W_i,\theta)] & = E_P[f_{j}(W_i)]  + g_{j}(\theta) \\
     & = E_P[f_{j}(W_i)]   +\bigg[ - G_r((-\infty,-x_1'(\beta_1 + \Delta_1) \times [-x_2'\beta_2,\infty)))p_x\bigg],
    \end{align*}\normalsize
where $x \in \mathcal{X}$ and $p_x$ is the probability of support point $x$ occurring. $G_r(\cdot,\cdot)$ is the Bivariate Gaussian process with correlation $r$.  \code{DGP=6} assumes $r=0$, so the moment $g_{j}(\theta)$ can be expressed as:
\[
g_{j}(\theta) =  -\Phi(-x_1'(\beta_1 + \Delta_1))(1- \Phi(-x_2'\beta_2))p_x,
\]
where $\Phi(\mu)$ is the univariate Gaussian CDF with mean $\mu$ and variance equal to $1$.  Compare to \code{Line 143} in \code{moments\_theta.m}:
\footnotesize
    \begin{lstlisting}[backgroundcolor = \color{gray!30},
                   xleftmargin = 0cm,
                   framexleftmargin = 1em]
g_ineq((ii-1)*2 + 1,1) = normcdf(-x1*(beta1+delta1))*(1-normcdf(-x2*beta2))*pX;
    \end{lstlisting}\normalsize
(The negative of \code{g\_ineq} is reported on \code{Line 164} in \code{moments\_theta.m} to get the correct sign.)

\subsubsection*{Step 4: Standard Deviation Estimator for $\sigma_j(W_i)$}
Under the assumption that the moment functions are separable, the standard deviation does not depend on $\theta$.  Specify the standard deviation estimator in the function \code{moments\_stdev}.  Override  \code{moments\_stdev.m} with the following shell:
\footnotesize
\begin{lstlisting}[backgroundcolor = \color{gray!30},
                   xleftmargin = 0cm,
                   framexleftmargin = 1em]
[f_stdev_ineq,f_stdev_eq] = moments_stdev(W,f_ineq,f_eq,J1,J2,KMSoptions)
    f_stdev_ineq            = zeros(J1,1);
    f_stdev_eq              = zeros(2*J2,1);
%% Define output here %%
end
\end{lstlisting} \normalsize
The inputs are: data \code{W}, data-implied moment functions \code{f\_ineq} and \code{f\_eq}, number of moment (in)equalities \code{J1} and \code{J2}, and a structure of options \code{KMSoptions}.  The outputs are the $J_1$-by-$1$ vector of standard deviations for the moment inequalities $\code{f\_stdev\_ineq}$ and the $2J_2$-by-$1$ vector of standard deviations for the moment equalities $\code{f\_stdev\_eq}$.

For  \code{DGP=6}, the estimator for $\sigma_j(W_i)$ is
\begin{align*}
\hat \sigma_{j} = \sqrt{ \frac{1}{n}\sum_{i=1}^n \mathbbm{1}(Y_i = y, X_i=x) \left(1 - \frac{1}{n}\sum_{i=1}^n \mathbbm{1}(Y_i = y_i, X=x)  \right)}
\end{align*}
Compare to \code{Lines 58-59} in \code{moments\_stdev.m}:
\footnotesize
    \begin{lstlisting}[backgroundcolor = \color{gray!30},
                   xleftmargin = 0cm,
                   framexleftmargin = 1em]
f_stdev_ineq(:,1) = sqrt(abs(f_ineq).*(1-abs(f_ineq)));
f_stdev_eq(:,1)   = sqrt(abs(f_eq).*(1-abs(f_eq)));
    \end{lstlisting}\normalsize

\subsubsection*{Step 5: Gradient of Model-implied Moment Function $Dg_j(\theta)$}
The CPI algorithm requires that the user specifies gradients of the moment functions.  Since the moments are separable, the gradient does not depend on data $W$, so that:
\[
\frac{\partial m_j(W,\theta)}{\partial \theta_k} = \frac{\partial g_j(\theta)}{\partial \theta_k}.
\]
The gradients are specified in  the function \code{moments\_gradient.m}, and below is a shell for this function:
\footnotesize
\begin{lstlisting}[backgroundcolor = \color{gray!30},
                   xleftmargin = 0cm,
                   framexleftmargin = 1em]
[Dg_ineq,Dg_eq] = moments_gradient(theta,J1,J2,KMSoptions)
    dim_p         = KMSoptions.dim_p;
    Dg_ineq       = zeros(J1,dim_p);
    Dg_eq         = zeros(2*J2,dim_p);
%% Define output here %%
end
\end{lstlisting} \normalsize
The inputs are: a $d$-by-$1$ parameter vector $\code{theta}$, the number of moment (in)equalities \code{J1} and \code{J2}, and the structure of options \code{KMSoptions}.  The outputs are the $J_1$-by-$d$ matrix of gradients for the moment inequalities \code{Dg\_ineq}, where
\begin{align*}
\text{\code{Dg\_ineq}}_{j,k} = \frac{\partial g_j(\theta)}{\partial \theta_k} \quad j =1,\cdots,J_1,\quad  k =1,\cdots,d
\end{align*}
and the $2 J_2$-by-$d$ matrix of gradients for the moment equalities \code{Dg\_eq}, where
\begin{align*}
\text{\code{Dg\_eq}}_{j,k} = \frac{\partial g_j(\theta)}{\partial \theta_k} \quad j =J_1+1,\cdots,J.
\end{align*}
As an example, consider the moment inequality in Equation (5.3) \citeA{kaido2017confidence}:
\[
g_j(\theta) = - \Phi(-x_1'(\beta_1 + \Delta_1))(1- \Phi(-x_2'\beta_2))p_x.
\]
The gradients are:
\begin{align*}
\frac{\partial g_j(\theta)}{\partial \beta_1} & = x_1' \phi(-x_1'(\beta_1 + \Delta_1))(1-\Phi(-x_2'\beta_2))p_x \\
\frac{\partial g_j(\theta)}{\partial \beta_2} & = -x_2' \Phi(-x_1'(\beta_1 + \Delta_1))\phi(-x_2'\beta_2)p_x \\
\frac{\partial g_j(\theta)}{\partial \Delta_1} & = x_1' \phi(-x_1'(\beta_1 + \Delta_1))(1-\Phi(-x_2'\beta_2))p_x \\
\frac{\partial g_j(\theta)}{\partial \Delta_2} & =0,
\end{align*}
where $\phi(\mu)$ is the univariate Gaussian PDF with mean $\mu$ and variance equal to $1$. Compare to \code{Lines 112-115} in \code{moments\_gradient.m}:
\footnotesize
    \begin{lstlisting}[backgroundcolor = \color{gray!30},
                   xleftmargin = 0cm,
                   framexleftmargin = 1em]
Dg3b1 = x1.*normpdf(-x1*(beta1+delta1)) .*(1-normcdf(-x2*beta2))*pX;
Dg3b2 = -x2.*normcdf(-x1*(beta1+delta1)).*normpdf(-x2*beta2)*pX;
Dg3d1 = x1.*normpdf(-x1*(beta1+delta1)) .*(1-normcdf(-x2*beta2))*pX;
Dg3d2 = zeros(1,2);
    \end{lstlisting}\normalsize

\subsubsection*{Step 6: Algorithm Options}
Algorithm options are specified in the file \code{KMSoptions.m}.  These options should be adjusted for each user-specified model in order to balance computational time and accuracy.  The key options are highlighted below.
\begin{itemize}
\item \code{KMSoptions.parallel} turns on parallel computing if set equal to $1$ (\code{Line 35}).
\item \code{KMSoptions.CVXGEN} uses \code{CVXGEN} if set equal to $1$ (\code{Line 38}).
\item \code{KMSoptions.B} specifies the number of bootstrap repetitions (\code{Line 42}).
\item \code{KMSoptions.EAM\_maxit} specifies the maximum number of EAM iterations (\code{Line 43}).
\item \code{KMSoptions.mbase} sets the base-multiplier for the initial number of points in the EAM algorithm (\code{Line 44}).  In order to get a better initial approximating surface increase this number.  There is a trade off between allowing the EAM algorithm to better approximate the surface near the global maximizer and obtaining a good initial fit.
\item \code{KMSoptions.h\_rate} determines the rate at which the parameter space is contracted (\code{Line 45}).  Set equal to a number between $1$ and $2$.  See the M-step in Section \ref{sec:EAM} on Page 25.
\item   \code{KMSoptions.h\_rate2} should be set to a number between $1$ and \code{KMSoptions.h\_rate} (\code{Line 46}).  See the Section \ref{sec:EAM}.
\item \code{KMSoptions.EAM\_obj\_tol} is one of the convergence criteria (\code{Line 48}).  It is required that the absolute difference between the expected improvement maximizer value and $p'\theta^{*,L}$ is less than or equal to \code{KMSoptions.EAM\_obj\_tol}.
\item \code{KMSoptions.EAM\_maxviol\_tol} is another convergence criterion (\code{Line 56}). It demands that the maximum moment violation is close to $0.$  Set equal to \code{inf} to turn off.
\item \code{KMSoptions.EI\_points} sets the minimum number of initial points with positive expected improvement for the $M$ step \code{Line 132}.
\end{itemize}

\subsubsection*{Step 7: Determine Input Parameters}
The following inputs are required:
\begin{enumerate}
\item \code{W} is an $n$-by-$d_W$ matrix of data.  $n$ is the number of observations and $d_W$ is the number of variables in $W$.  In our working example, $n= 4000$ is the number of markets and $d_W = 6$.  The first and second variables in $W$ are the entry decisions of firms $1$ and $2$.  The fourth and sixth variables in $W$ are the random market characteristics of firms $1$ and $2$.  The third and fifth variables in $W$ are constants.  The market characteristic has support $\{-1,1\}$.  See \code{Lines 41-111} in \code{ExampleDGP6.m} for more detail on how the data is generated.

\item \code{theta\_0} is a $d$-by-$1$ vector.  It is the initial guess for the parameter vector $\theta$.  We arbitrarily set  \code{theta\_0} to be the midpoint in the hyperrectangle $\{\theta \in \mathbb{R}^d : \theta_{LB} \leq \theta \leq \theta_{UB}\}$ in the working example.  See \code{Line 29} in \code{ExampleDGP6.m}.  Valid input is any value in the parameter space.

\item \code{p} is the $d$-by-$1$ directional vector in the problem $\sup_{\theta \in \Theta} p'\theta$ subject to $\sqrt{n} \frac{ \bar m_j(\theta)}{\sigma_j} \leq \hat c(\theta)$.  Valid input is $p_i = 1$ for any one component and $0$ otherwise.  See \code{Lines 30-31} in \code{ExampleDGP6.m}.  See Appendix \ref{appendix:polytope} for non-basis directional vectors.

\item \code{theta\_feas} is a $K$-by-$d$ matrix of $K$ feasible $\theta$ stacked in row format:
\[
\begin{bmatrix} \theta_1' \\ \vdots \\ \theta_K'\end{bmatrix}.
\]
A feasible $\theta$ is one that satisfies $\sqrt{n} \frac{\bar m_j(\theta)}{\sigma_j} \leq \hat c(\theta), \forall j=1,\cdots,J$.  This input is optional, and if set to the empty set,  the algorithm attempts to find a feasible point using an auxiliary search. $\code{theta\_feas}$ is set equal to \code{[]} on \code{Line 115} in \code{ExampleDGP6.m}.

\item \code{LB\_theta} and \code{UB\_theta} are $d$-by-$1$ vectors defining the lower and upper bounds of the parameter space.  Valid input is $\theta_{LB} \leq \theta_{UB}$. See \code{Lines 27-28} in \code{ExampleDGP6.m}.

\item  \code{A\_theta} and \code{b\_theta} are $L$-by-$d$ and $L$-by-$1$ matrices defining the polytope constraints on the parameter space.  It is required that the parameter space has a non-empty interior.  \code{A\_theta} and \code{b\_theta} are set to \code{[]}, see \code{Line 115} in \code{ExampleDGP6.m}.

\item \code{alpha} is the nominal significance level.  Valid input is a number in $[0,0.5]$.   It is set equal to $0.05$, see \code{Line 6} in \code{ExampleDGP6.m}.

\item \code{type} determines if either a two-sided or one-sided confidence interval is computed. Valid input is either \code{'two-sided'}  or  \code{'one-sided'}. \code{type} is set equal to \code{'two-sided'} for a two-sided confidence interval, see \code{Line 21} in \code{ExampleDGP6.m}.

 \item \code{method} determines whether a Calibrated Projection Interval or an AS Projection Interval is computed.  Valid input is either  \code{'KMS'}  or  \code{'AS'}.  \code{method} is set equal to \code{'KMS'}, see \code{Line 4} in \code{ExampleDGP6.m}.

 \item \code{kappa} specifies the tuning parameter $\kappa$.  Valid input is either  \code{NaN} for the default $\kappa = \sqrt{\ln(n)}$ or a user-specified function \code{@(n)kappa\_function(n)} satisfying  Assumption 4.2 in \citeA{kaido2017confidence}.   \code{kappa} is set equal to \code{NaN}, see \code{Line 22} in \code{ExampleDGP6.m}.

 \item \code{phi} specifies the GMS function $\varphi_j(x)$.  Valid input is either  \code{NaN} for the default hard thresholding function
     \[
     \varphi_j(x) = \begin{cases} 0 & \text{if } x \geq -1 \\ -\infty & \text{else} \end{cases}
      \]
      or a user-specified function \code{@(x)phi\_function} satisfying Assumption 4.2.   \code{phi} is set equal to \code{NaN}, see \code{Line 23} in \code{ExampleDGP6.m}.

\item \code{CVXGEN\_name} is the name for the \code{CVXGEN MEX} file, see Section \ref{subsec:cvxgen}.   \code{CVXGEN\_name}  is set equal to \code{'ExampleDGP6'}, see \code{Line 39} in \code{ExampleDGP6.m}.

\item \code{KMSoptions} is the structure of algorithm options.  It is called on \code{Line 9} in \code{ExampleDGP6.m} and updated further throughout the file.
\end{enumerate}

Finally, the CPI algorithm is called on \code{Lines 114-115} in \code{ExampleDGP6.m}, with the inputs specified below:
\footnotesize
\begin{lstlisting}[backgroundcolor = \color{gray!30},
                   xleftmargin = 0cm,
                   framexleftmargin = 1em]
[KMS_confidence_interval,KMS_output] = KMS_0_Main(W,theta_0, p,[],LB_theta,...
   UB_theta,[],[],alpha,type,method,kappa,phi,CVXGEN_name,KMSoptions);
\end{lstlisting} \normalsize

\section{Calibrated Projection Interval Algorithm} \label{sec:algorithm}
In this section we provide an overview of the CPI algorithm.  We start with \code{KMS\_0\_Main}, since this is the function that calls the CPI algorithm.  The empirical moments $\hat f_j$, standard deviation $\hat \sigma_j$, and the recentered bootstrap moments $\mathbb{G}_j^b$ do not depend on $\theta$ and can be computed outside of the EAM algorithm.  These are computed on \code{Lines 217, 502}, and \code{Lines 438-493} and \code{509-510} in \code{KMS\_0\_Main}, respectively.  The recentered bootstrap moments are denoted \code{G\_ineq} and \code{G\_eq}.

It is required that the EAM algorithm is initiated with a feasible point, that is, a point $\theta^{\text{feas}} \in \Theta$ satisfying:
\begin{align*}
\sqrt{n} \frac{\hat f_j + g_j(\theta^{\text{feas}})}{\hat \sigma_j} \leq \hat c(\theta^{\text{feas}}) \quad \forall j=1,\cdots,J.
\end{align*}
The CPI algorithm executes two feasible search algorithms on \code{Lines 565-583}  in \code{KMS\_0\_Main}.  The feasible search algorithms are \code{KMS\_1\_FeasibleSearch.m} and \code{KMS\_2\_EAM\_FeasibleSearch.m}.  If a feasible point(s) is supplied by the user, then the algorithm skips this step.

The EAM algorithm is called on \code{Lines 602-635} in \code{KMS\_0\_Main}.  The search direction $p$ is executed first and the search direction $-p$ second.  Output including the optimal point $\theta^{*,EAM}$, the calibrated critical value at this point $\hat c(\theta^{*,EAM})$, the expected improvement $EI(\theta^{*,EAM})$, and the convergence time is reported in the structure $\code{KMS\_output}$.

An optional algorithm on \code{Lines 637-676} in \code{KMS\_0\_Main} is also included  (set \\ \code{KMSoptions.direct\_solve=1} to run this algorithm).  This algorithm solves the problem:
\begin{align*}
& \text{min/max}_{\theta \in \Theta} p'\theta \\
& \text{s.t. } \sqrt{n} \frac{\hat f_j + g_j(\theta)}{\hat \sigma_j} \leq \hat c(\theta) \quad \forall j=1,\cdots,J
\end{align*}
using numerical gradients (the gradient of $\hat c(\theta)$ is unknown).  Even for simple problems this algorithm requires a large amount of computational time to find a solution and the solution is often not the global minimizer/maximizer.\footnote{One could specify an analytical gradient function, where the analytical gradients $D_{\theta}g_j(\theta)$ are passed and the numerical gradient for $\hat c(\theta)$ is computed.  This is not done in this version of the CPI algorithm.}

Finally, the Calibrated Projection Interval (or AS Projection Interval) is reported on \code{Lines 678-707} in \code{KMS\_0\_Main}.  The feasible search and EAM algorithm is discussed in Sections \ref{sec:feassearch} and \ref{sec:EAM} below. The algorithm that computes $\hat c(\theta)$ is discussed in Section \ref{sec:critval}

\subsection{Feasible Search Algorithm}\label{sec:feassearch}
The feasible search algorithms \code{KMS\_1\_FeasibleSearch} and \code{KMS\_2\_EAM\_FeasibleSearch}  attempt to find a point $\theta$ satisfying:
\begin{align}\label{eq:feas}
\theta \in \Theta^{\text{feas}} \equiv \left\{ \theta \in \Theta : \sqrt{n} \frac{\hat f_j + g_j(\theta^{\text{feas}})}{\hat \sigma_j} \leq \hat c(\theta^{\text{feas}}) \quad \forall j=1,\cdots,J\right\}.
\end{align}
The algorithm \code{KMS\_1\_FeasibleSearch} solves the problem:
\begin{align}\label{eq:feas1}
\min_{\theta \in \Theta} \max_{j=1,\cdots,J}\sqrt{n} \frac{\hat f_j + g(\theta)}{\hat \sigma_j}.
\end{align}
Let the minimizer be $\theta^{*,FS1}$.  The hope is that $\theta^{*,FS1}$  satisfies the relaxed condition
\begin{align}\label{eq:constraint}
\sqrt{n} \frac{\hat f_j + g_j(\theta^{*,FS1})}{\hat \sigma_j} \leq \hat c(\theta^{*,FS1}), \forall j=1,\cdots,J.
\end{align} If $\max_j \sqrt{n} \frac{\hat f_j + g_j(\theta^{*,FS1})}{\hat \sigma_j} \leq 0$, then the condition in Equation \eqref{eq:constraint} is satisfied since $\hat c(\theta) \geq 0, \forall \theta \in \Theta$.

A MultiStart algorithm is used to solve Problem \eqref{eq:feas1}.\footnote{The solver \code{fmincon} cannot efficiently solve Problem \eqref{eq:feas1} as written, since the gradient of $ \max_{j=1,\cdots,J}\sqrt{n} \frac{\hat f_j + g(\theta)}{\hat \sigma_j}$ is unknown.  By introducing a free parameter $\gamma$, the problem can be re-written so that \code{fmincon} can solve it.  See Section \ref{sec:EAM}.}   On \code{Lines 72-78} in \code{KMS\_1\_Feasible\-Search}, a set of starting points is drawn uniformly from $\Theta$.  These starting points are passed to \code{fmincon} (\code{Lines 93-119}).  At each solution, the constraint violation is computed on \code{Line 134}.  If there is a feasible point (constraint violation = 0), then the feasible point \code{theta\_feas} is returned.  Otherwise, \code{flag\_feas=0} is returned indicating failure to find a feasible point.

If the feasible search algorithm \code{KMS\_1\_FeasibleSearch} fails, the second feasible search algorithm, \code{KMS\_2\_EAM\_FeasibleSearch}, is executed.  It uses an EAM-type algorithm to try to find a feasible point.  In particular, \code{Line 86-92} in \code{KMS\_2\_EAM\_FeasibleSearch.m} draws an initial set of points $\theta^{(1)},\cdots,\theta^{(L)}$.  The calibrated critical value $\hat c(\theta)$ is computed at each of these points (E-step, \code{Line 109}).  If any of these points are feasible, the algorithm is terminated and a feasible point is returned (\code{Line 125-137}).  Otherwise, the surface $c_L(\theta)$ is approximated using the kriging method (A-step, \code{Lines 139-144}).  Last, the following minimization problem is solved using a MultiStart algorithm:
\begin{align}\label{eq:feas2}
\min_{\theta \in \Theta} \max_{j=1,\cdots,J} \left( \sqrt{n} \frac{\hat f_j + g(\theta)}{\hat \sigma_j} - c_L(\theta)\right),
\end{align}
(M-step, \code{Lines 174-204}).  Call the minimizer $\theta^{*,L+1}$.  This EAM algorithm is re-iterated with the new set of points $\{\theta^{(l)}\}_{l=1}^L\cup \{\theta^{*,L+1}\}$, and continues until either a feasible point is obtained or the maximum number of iterations \code{KMSoptions.EAM\_maxit} is reached.

\subsection{EAM Algorithm}\label{sec:EAM}
The EAM algorithm (\code{KMS\_3\_EAM}) for search direction $q \in \{-p,p\}$ is called on \code{Lines 603-635} in \code{KMS\_0\_Main.m}.  The inputs for the EAM algorithm are:
\begin{enumerate}
\item \code{q} is the directional vector, set equal to either $p$ or $-p$.

\item \code{sgn\_q} is equal to $-1$ or $1$.  It specifies whether we are maximizing in direction $q=p$ or $q=-p$.

\item \code{theta\_feas} is a $K$-by-$d$ matrix of feasible points stacked in row format.

\item \code{theta\_init}, \code{c\_init},  \code{CV\_init}, \code{maxviol\_init} are a set of $\theta$s, calibrated critical values,  constraint violations, and maximum violations passed from the feasible search algorithm.  These can be empty.

\item \code{f\_ineq}, \code{f\_eq}, \code{f\_ineq\_keep}, \code{f\_eq\_keep} are output from \code{moments\_w.m}, which are the data-implied moment (in)equalities and the moment (in)equalities that we keep.

\item \code{f\_stdev\_ineq} and \code{f\_stdev\_eq} are output from \code{moments\_stdev.m}, which are the standard deviations of the moment (in)equalities.

\item \code{G\_ineq} and \code{G\_eq} are the recentered bootstrap moment (in)equalities.

\item \code{KMSoptions} is a structure of algorithm options.
\end{enumerate}
The outputs are:
\begin{enumerate}
\item \code{theta\_hat} is the $d$-by-$1$ solution to $\max_{\theta \in \Theta} q'\theta$ subject to the calibrated moment inequality constraints from the EAM algorithm.  Call this optimal point $\theta^{*,EAM}$.

\item \code{theta\_optbound} is the value $q'\theta^{*,EAM}$.

\item \code{c} is the calibrated critical value $\hat c(\theta^{*,EAM})$.

\item \code{CV} is the maximum constraint violation at $\theta^{*,EAM}$.  This should be zero if $\theta^{*,EAM}$ is feasible.

\item \code{EI} is the expected improvement at $\theta^{*,EAM}$.

\item \code{flag\_opt} is a flag equal to $1$ if the EAM algorithm converged.
\end{enumerate}
The key steps in the EAM algorithm are detailed below.

\textbf{Initialization}:
A initial set of points is drawn on \code{Lines 93-99} in \code{KMS\_3\_EAM}.  The set of feasible points \code{theta\_feas} is also added to the pool of initial points $\{\theta^{(l)}\}_{l=1}^{L_0}$.  The set of points to be evaluated is saved in the $L$-by-$d$ matrix \code{theta\_Estep} (where $L = L_0$ on the first iteration).

The following steps are iterated until either the program converges or a preset maximum number of iterations is reached.

\textbf{E-step, Evaluation}:
\code{Line 143} in \code{KMS\_3\_EAM} calls the function  \code{KMS\_31\_Estep}. Within this function, \code{c\_Estep} is the $L$-by-$1$ vector of calibrated (or AS) critical values $\hat c(\theta)$ for each $\theta \in \{\theta^{(l)}\}_{l=1}^{L}$;  \code{CV\_Estep} and \code{maxviol\_Estep} are the constraint violation and maximum violation for each $\theta \in \{\theta^{(l)}\}_{l=1}^{L}$, respectively.  The subfunctions \code{KMS\_31\_Estep}, \code{KMS\_32\_Critval}, and  \code{KMS\_33\_Coverage} are discussed in more detail in Section \ref{sec:critval}.

\code{Lines 145-155} in \code{KMS\_3\_EAM} prepare the matrices \code{theta\_Astep} and \code{c\_Astep}, which keep track of all points to be passed to the A-step.  The kriging model is sensitive if two points $\theta^{(l)}$ and $\theta^{(k)}$ are too close together.  Therefore, if two points are too close to one another, only one point is passed to the kriging model, see \code{Line 161} in \code{KMS\_3\_EAM}.

\textbf{A-step, Approximation:}
The surface $\hat c(\theta)$ is approximated via a kriging model.  The set of points $ \{\theta^{(l)},\hat c(\theta^{(l)})\}_{l=1}^{L}$ is passed to the kriging function on \code{Line 163} in \code{KMS\_3\_EAM}.  We use the \code{DACE} package \cite{lophaven2002dace}.  The \code{DACE MATLAB} files are saved in the subfolder \code{\textbackslash dace}.  The output is the structure \code{dmodel}.  The function \code{[c,Dc,mse,Dmse]=predictor(theta,dmodel)} uses the interpolated surface to predict the value of $\hat c(\theta)$ and gradient $D_{\theta} \hat c(\theta)$ at $\theta$.  The standard deviation $\hat \zeta s_L(\theta)$ is also estimated and is equal to $\sqrt{\text{\code{mse}}}$.\footnote{For more information and source files for the \code{DACE MATLAB} package, go to \url{http://www2.imm.dtu.dk/projects/dace/}.}

\textbf{M-step, Maximization:}  Using the approximated surface $c_L(\theta)$ and standard deviation $\hat \zeta s_L(\theta)$, the next point in the sequence $\{\theta^{(l)}\}_{l=1}^{L}$ is chosen to maximize the expected improvement function:
\begin{align}\label{eq:Mstep}
\theta^{*,\text{Mstep}} =\argmax_{\theta \in \Theta} \mathbb{E}\mathbb{I}(\theta) = (q' \theta - q' \theta^{*,L})_+ \left(1 - \Phi \left( \frac{ \max_{j=1,\cdots,J} \hat h_j(\theta) - c_L(\theta)}{\hat \zeta s_L(\theta)} \right) \right),
\end{align}
where
\begin{align*}
 \theta^{*,L}& \equiv \argmax_{\theta \in \{\theta^{(l)}\}_{l=1}^L}  q'\theta \quad \text{s.t.}  \quad \sqrt{n} \frac{\hat f_j + g_j(\theta)}{\hat \sigma_j} \leq \hat c(\theta), \quad  \forall j = 1,\cdots,J,
\\
(x)_{+} &\equiv \max(0,x),
\\
\hat h_j(\theta) & \equiv \sqrt{n} \frac{\hat f_j+ g_j(\theta)}{\hat \sigma_j}.
\end{align*}

Problem \eqref{eq:Mstep} is solved using a MultiStart algorithm.  Three important steps are required to resolve numerical issues with Problem \eqref{eq:Mstep}:
\begin{enumerate}
\item The gradients of the functions $(x)_{+}$ and $\max_{j=1,\cdots,J} \hat h_j(\theta)$ are undefined at crossing points (e.g, at $x=0$ for $(x)_+$).  Problem \eqref{eq:Mstep} is rewritten as:
     \[
        - \min_{\theta \in \Theta,q'\theta \geq q' \theta^{*,L}} \max_{j=1,\cdots,J}\left[-(q' \theta - q' \theta^{*,L})\Phi \left(- \frac{ \hat h_j(\theta) - c_L(\theta)}{\hat \zeta s_L(\theta)} \right)\right].
     \]
    The $\min/\max$ problem can be solved using $\code{fmincon}$ by introducing a free parameter $\gamma$ and rewriting the problem as a minimization problem with nonlinear constraints.
    \begin{align}\label{eq:Mstep2}
        & \min_{\gamma \in \mathbb{R}, \theta \in \Theta,q'\theta \geq q' \theta^{*,L}}  \gamma \\
        & \text{s.t.} \quad
        -(q' \theta - q' \theta^{*,L})\Phi \left(- \frac{  \hat h_j(\theta) - c_L(\theta)}{\hat \zeta s_L(\theta)} \right) - \gamma \leq 0, \forall j = 1,\cdots,J.\nonumber
    \end{align}
    The gradients of the objective function and constraints in Problem \eqref{eq:Mstep2} are well-defined, so it can be solved using \code{fmincon} with analytical gradients.  See the functions \code{KMS\_34\_EI\-\_objective.m} and \code{KMS\_35\_EI\_constraint.m}.  The same technique is used to solve Problems \eqref{eq:feas1} and \eqref{eq:feas2} in the feasible search algorithms.

\item Problem \eqref{eq:Mstep} can become ill-conditioned in the sense that the objective function is numerically equal to zero for $\theta$ such that the (in)equalities are modestly violated. Consequently, the non-linear programming solver \code{fmincon} may get stuck.  To overcome this issue, we use an auxiliary method to draw $\theta$ such that $\mathbb{E}\mathbb{I}(\theta)>0$ (see \code{Line 287} in \code{KMS\_3\_EAM} and see \code{KMS\_36\_drawpoints}).  These points are passed to \code{fmincon} and Problem \eqref{eq:Mstep2} is solved using MultiStart.

     To further explain the numerical issue, observe that the argument of $\Phi(\cdot)$ is
     \[
     - \frac{ \hat  h_j(\theta) - c_L(\theta)}{\hat \zeta s_L(\theta)}.
     \]
     If the approximated moment condition $\hat h_j(\theta) - c_L(\theta)$ is violated and hence positive, the argument of $\Phi(\cdot)$ is negative.  If, in addition, $\hat \zeta s_L(\theta)$ is small relative to $\hat h_j(\theta) - c_L(\theta)$, then the term $\Phi \left(  - \frac{  \hat h_j(\theta) - c_L(\theta)}{\hat \zeta s_L(\theta)}\right)$ can be numerically equal to zero.  Therefore, the objective function may have many local minima, and in applications \code{fmincon} may get stuck.

\item The expected improvement objective function trades off increasing the value of $q'\theta$ and increasing the likelihood that $\theta$ satisfies the moment conditions via the term $\Phi \left(- \frac{  \hat h_j(\theta) - c_L(\theta)}{\hat \zeta s_L(\theta)} \right)$. The approximation $c_L(\cdot)$ to $\hat c(\cdot)$ is not perfect.  Therefore, the value of $\Phi \left(- \frac{  \hat h_j(\theta) - c_L(\theta)}{\hat \zeta s_L(\theta)} \right)$ can be positive for $\theta \not \in \Theta^{\text{feas}}$ (defined in Equation \eqref{eq:feas}).  In later iterations of the EAM algorithm, we may find that the expected improvement maximizer $\theta^{*,\text{Mstep}}$ is not feasible.  Adding non-feasible $\theta^{*,\text{Mstep}}$ improves the fit of $c_L(\theta)$ and $\Phi \left(- \frac{  \hat h_j(\theta) - c_L(\theta)}{\hat \zeta s_L(\theta)} \right)$ converges in probability to $0$ for $\theta \not \in \Theta^{\text{feas}}$ as the number of EAM iterations increases.   To increase the likelihood that $\theta^{*,\text{Mstep}} \in \Theta^{\text{feas}}$, we contract the parameter space, which constrains the expected improvement function.  The contraction forces the expected improvement maximizer $\theta^{*,\text{Mstep}}$ to be near $\theta^{*,L}$ if the EAM algorithm begins to stall, increasing the likelihood that $\theta^{*,\text{Mstep}}$ is feasible.

    The contracted parameter space is:
    \begin{align}\label{eq:contract}
    \Theta(h_{\text{rate}}^{\text{counter}}) = \left\{\theta : \theta_{LB} \leq \theta \leq \theta_{UB}, A_{\theta} \theta \leq b_{\theta},  q'\theta^{*,L}  \leq q'\theta \leq q'\theta^{*,L}  + \frac{ q'\theta^{\dagger}- q'\theta^{*,L}}{h_{\text{rate}}^{\text{counter}}}\right\},
    \end{align}
     where $q'\theta^{\dagger} = \max_{\theta \in \Theta} q'\theta$.  For example, if $q=p$, then  $q'\theta^{\dagger} = q' \theta_{UB}$.  The term $h_{\text{rate}}^{\text{counter}}$ controls the rate of contraction.  $h_{\text{rate}}$ is specified by the user and is equal to \code{KMSoptions.h\_rate}.  The exponent \code{counter} is a natural number that increases by one on iteration $i$ of the EAM algorithm if insufficient progress is made on iteration $i-1$.  If $\theta^{*,L}$ is too close to the contracted boundary, then \code{counter} decreases by one.  Problem \eqref{eq:Mstep2} is solved subject to $\theta \in \Theta(h_{\text{rate}}^{\text{counter}})$.  Our convergence criteria (described below) are chosen to make sure that no mechanical convergence occurs.
\end{enumerate}
The contraction of the parameter space occurs on \code{Lines 200-273} in \code{KMS\_3\_EAM},  \code{Lines 281-328} draw points, and \code{Lines 336-373} call \code{fmincon}.

\textbf{Updating and Convergence:}
The final step in the EAM algorithm is to update the set of points and check convergence criteria.  Provided we find a point with positive expected improvement in the M-step, we add both the M-step solution  as well as a uniformly-drawn point to the set of evaluation points $\{\theta^{(l)}\}_{l=1}^L$. We also add two points, $\theta_{\epsilon_1}$ and $\theta_{\epsilon_2}$ that are close to $\theta^{*,L}$ and satisfy $q'\theta_{\epsilon_k} > q' \theta^{*,L}$.  The distance between $\theta_{\epsilon_k}$ and $\theta^{*,L}$  is determined by the option \code{KMSoptions.h\_rate2}.  See \code{Lines 430-439} in \code{KMS\_3\_EAM}.

The convergence check occurs on \code{Lines 464-485} in \code{KMS\_3\_EAM}.  We first check if $\theta^{*,L}$ is too close to the boundary of $\Theta$.  If $|p' \theta^{*,L} - p' \theta_{UB}| < 10^{-4}$ (for search direction $p$), then a warning that the parameter is on the boundary is displayed and we output $\theta^{*,\text{EAM}} = \theta^{*,L}$.  Otherwise, if all of the following conditions are satisfied, we say that the EAM algorithm has converged, and we output $\theta^{*,\text{EAM}} = \theta^{*,L}$.
\begin{enumerate}
\item \code{iter >= EAM\_minit}: The current iteration $i$ of the EAM algorithm is greater than or equal to \code{KMSoptions.EAM\_minit}.  This ensures that the EAM algorithm does not terminate early.  Default is \code{KMSoptions.EAM\_minit=4}.

\item \code{change\_EI\_proj < EAM\_obj\_tol}: The absolute difference in the value of the objective function $q'\theta$ between the expected improvement maximizer $\theta^{*,\text{Mstep}}$ and the current feasible optimal $\theta^{*,L}$ is less than  the tolerance parameter \code{KMSoptions.EAM\_obj\_tol}.

\item \code{change\_proj < EAM\_tol}: The absolute difference in the value of the objective function $q'\theta$ between the current feasible optimal $\theta^{*,L}$ and the previous iteration's feasible optimal $\theta^{*,L-1}$ is less than  the tolerance parameter \code{KMSoptions.EAM\_toll}.

\item \code{feas\_points>num\_feas}: We have found at least one feasible point inside the EAM algorithm.  This ensures that we do not terminate the EAM algorithm with only the feasible points that are fed into the EAM function.

\item \code{abs(opt\_dagger - q.'*theta\_hash) > 1e-4}: The point $\theta^{*,L}$ is not too close to the boundary of the contracted parameter space.  This ensures that the EAM algorithm is not terminated mechanically.

\item \code{abs(maxviol\_hash) <EAM\_maxviol\_tol}: The value $\bigg| \max_{j=1,\cdots,J} \sqrt{n} \frac{\hat f_j + g_j(\theta^{*,L})}{\hat \sigma_j} - \hat c(\theta^{*,L})\bigg|$ is less than the tolerance parameter \code{KMSoptions.EAM\_maxviol\_to}.  If this condition is violated, then by continuity of $\hat c(\cdot)$ it is possible to increase the value of $q'\theta^{*,L}$ and not violate the moment conditions.
\end{enumerate}

Last, the contraction counter is updated (\code{Lines 487-494}).  If $|q'\theta^{*,L} - q'\theta^{*,L-1}| < 1 \times 10^{-6}$ (so that no progress is made between this iteration and the previous iteration), then the contraction counter is increased by one: \code{counter = counter + 1}.  If the contraction counter is positive and $\theta^{*,L}$ is too close to the contracted parameter space the contraction counter is decreased by one: \code{counter = counter - 1}.

\subsection{Root-Finding Algorithm Used to Compute $\hat{c}_n(\theta)$}\label{sec:critval}
This section explains in detail how to compute the calibrated critical value $\hat{c}(\theta)$:
\begin{align}
\hat c(\theta) \equiv \inf \{c \in \mathbb{R}_+ : P^*(\Lambda^b(\theta,\rho,c) \cap p'\lambda = 0 \} \neq \emptyset \geq 1-\alpha\},
\end{align}
where $P^*$ is the bootstrap empirical frequency.  The relevant \code{MATLAB} files are: \code{KMS\_31\_Estep.m}, \code{KMS\_32\_Critval.m}, and  \code{KMS\_33\_Coverage.m}.  For a given $\theta \in \Theta$, $P^*(\Lambda^b (\theta ,\rho ,c)\cap \{p^{\prime }\lambda =0\}\neq \emptyset)$ increases in $c$ (with $\Lambda^b (\theta ,\rho ,c)$ defined in Equation \eqref{eq:lambda}), and so $\hat c(\theta)$ can be quickly computed via a root-finding algorithm, such as the Brent-Dekker Method (BDM), see \citeA{brent1971algorithm} and \citeA{dekker1969finding}.  To do so, define $\Psi_{\alpha}(c) = \frac{1}{B} \sum_{b=1}^B \psi_b(c) - (1-\alpha)$ where
\[
\psi_b(c(\theta)) = \mathbf{1}(\Lambda^b(\theta,\rho,c) \cap \{p'\lambda = 0\}\neq \emptyset).
\]
Let $\bar c(\theta)$ be an upper bound on $\hat c(\theta)$, for example, the asymptotic Bonferroni bound $\bar c(\theta) \equiv \Phi^{-1}(1-\alpha/J)$.  It remains to find $\hat c(\theta)$ such that $\Psi_{\alpha}(\hat c(\theta)) =  0$ when $\Psi_{\alpha}(0)\leq 0$.  It is possible that $\Psi_{\alpha}(0) > 0$ in which case we output $\hat c(\theta) = 0$.  Otherwise, we use BDM to find the unique root to $\Psi_{\alpha}(c)$ on $[0,\bar c(\theta)]$ where, by construction, $\Psi_{\alpha}(\bar c(\theta)) \geq 0$.  We propose the following algorithm:

\textbf{Step 0} (Initialize)
\begin{enumerate}
\item Set $\mathit{Tol}$ equal to a chosen tolerance value;
\item Set $c_L = 0$ and $c_U =\bar c(\theta)$ (values of $c$ that bracket the root $\hat c_n(\theta)$);
\item Set $c_{-1} = c_L$ and $c_{-2} = \bot$  (proposed values of $c$ from $1$ and $2$ iterations prior).  Also set $c_0 = c_L$ and $c_1 = c_U$.
\item Compute $\varphi_j( \hat \xi_{j}(\theta))$ for $j = 1,\cdots,J$;
\item Compute $D_{\theta}g_j(\theta)$ for $j=1,\cdots,J$;
\item Compute $\mathbb{G}^b_{j}$ for $b = 1,\cdots,B$, $j = 1,\cdots,J$;
\item Compute $\psi_b(c_L)$ and $\psi_b(c_U)$ for $b = 1,\cdots,B$;
\item Compute $\Psi_{\alpha}(c_L)$ and $\Psi_{\alpha}(c_U)$.
\end{enumerate}

\textbf{Step 1} (Method Selection)

Use the BDM rule to select the updated value of $c$, say $c_2$.  The value is updated using one of three methods: Inverse Quadratic Interpolation, Secant, or Bisection.  The selection rule is based on the values of $c_{i}$, $i=-2,-1,0,1$ and the corresponding function values.

\textbf{Step 2} (Update Value Function)

Update the value of $\Psi_{\alpha}(c_2)$.  We can exploit the previous computation and monotonicity of the function $\psi_b(c_2)$ to reduce computational time:
\begin{enumerate}%\setlength{\itemsep}{-10pt}
\item If $\psi_b(c_L) = \psi_b(c_U) = 0$, then $\psi_b(c_2) = 0$;
\item If $\psi_b(c_L) = \psi_b(c_U) = 1$, then $\psi_b(c_2) = 1$.
\end{enumerate}

\textbf{Step 3} (Update)
\begin{enumerate}
\item If $\Psi_{\alpha}(c_2) \geq 0$, then set $c_U = c_2$.  Otherwise set $c_L = c_2$.
\item Set $c_{-2} = c_{-1}$, $c_{-1} = c_0$, $c_0 = c_L$, and $c_1 = c_U$.
\item Update corresponding function values $\Psi_{\alpha}(\cdot)$.
\end{enumerate}

\textbf{Step 4} (Convergence)
\begin{enumerate}
\item If $\Psi_{\alpha}(c_U) \leq \mathit{Tol}$ or if $|c_U - c_L| \leq \mathit{Tol}$, then output $\hat c_n(\theta) = c_U$ and exit. Note: $\Psi_{\alpha}(c_U) \geq 0$, so this criterion ensures that we have \emph{at least} $1-\alpha$ coverage.
\item Otherwise, return to \textbf{Step 1}.
\end{enumerate}
The computationally difficult part of the algorithm is computing $\psi_b(\cdot)$ in \textbf{Step 2}.  This is simplified for two reasons.  First, evaluation of $\psi_b(c)$ entails determining whether a constraint set comprised of $J+2d+2$ linear inequalities in $d$ variables is feasible.  This can be accomplished by efficiently employing \code{CVXGEN} or \code{CVX}.  Second, we exploit monotonicity in $\psi_b(\cdot)$ to reduce the number of linear programs needed to be solved.

The file \code{KMS\_31\_Estep} fixes $\theta$ and computes the GMS function $\varphi(\hat \xi_j(\theta))$ and gradients $D_{\theta}g_j(\theta)$;   \code{KMS\_32\_Critval} executes the BDM algorithm; and \code{KMS\_33\_Coverage} computes $\psi_b(c)$ for $b=1,\cdots,B$.

\section{Discussion} \label{sec:conclusion}

We have described how to implement the CPI algorithm to solve
\begin{align}\label{eq:dis}
& \inf/\sup_{\theta \in \Theta} p'\theta \\
& \text{s.t.} \sqrt{n} \frac{\bar m_{j}(\theta)}{\hat \sigma_{j}(\theta)} \leq \hat c(\theta) \quad j=1,\cdots,J.
\end{align}
One difficulty in solving this problem is that $\hat c(\theta)$ is a ``black-box function'' with an unknown gradient.  Directly solving this problem using \code{fmincon} with numerical gradients is slow and can return local solutions that are far from the global solution(s). The EAM algorithm is employed to solve this problem. This manual and \code{MATLAB} package can serve as a guide on how to implement the EAM algorithm to solve other black-box functions, provided sufficient continuity assumptions hold.

In the next release of the \code{MATLAB} package, we will allow for non-separability of the moment (in)equalities.  Additional numerical issues are presented in this case.  The estimator for the moment function, the estimator for the standard deviation, and the bootstrap need to be recomputed at each visit of $\theta \in \Theta$ in the EAM algorithm.  The gradients of the moment functions also depend on the data, so this adds additional numerical complexity in the optimization routine.

Another feature we plan to incorporate is an objective function $h(\theta)$ in Equation \eqref{eq:dis} that is not necessarily equal to $p'\theta$. The objective function, for example, could be a welfare function from a partially identified model that is parameterized by $\theta$ (for example, see \citeA{Barseghyan2017}).  Consequently, the welfare function is also partially identified.  Using the CPI algorithm one can obtain uniformly valid bounds on the function $h(\theta)$.  Modifications to the feasible search, fixed-point algorithm, and M-step are required.

\bibliographystyle{apacite}
\begin{flushleft}
\bibliography{library}
\end{flushleft}

\clearpage

\begin{appendices}

\section{Tables} \label{sec:tab}

\begin{table}[h!]
\centering
\begin{tabular}{llllll}
\hline
\code{DGP} & \code{dim\_p} & \code{J1} & \code{J2} & \code{S} & \code{CVXGEN\_name} \\ \hline
1-3          & 2               & 4           & 0           & 0          & \code{csolve\_DGP1} \\
4            & 2               & 8           & 0           & 0          & \code{csolve\_DGP4} \\
5-6          & 8               & 8           & 8           & 0          & \code{csolve\_DGP5} \\
7            & 9               & 8           & 8           & 0          & \code{csolve\_DGP7} \\
8            & 5               & 8           & 4           & 13         & \code{csolve\_DGP8} \\ \hline
\end{tabular}
\caption{List of parameters for creating the \code{CVXGEN MEX} files for simulations in \protect\citeA{kaido2017confidence}.  The first column corresponds to the parameter \code{DGP} in \code{KMS\_Simulation}.}
\label{tab:cvxgenparams}
\end{table}

\section{Polytope Constraints and Non-basis Directional Vectors} \label{appendix:polytope}
In this appendix we describe the numerical issues that arise when either $p$ is a non-basis directional vector or polytope constraints are imposed on the parameter space. We also propose a method on how to resolve these issues.  The key issue is how to draw points from the contracted parameter space, see Equation \eqref{eq:contract}. If the constraints $A_{\theta} \theta \leq b_{\theta}$ are included or if  $p$ is not a basis vector, then the contracted parameter space is a polytope but not a hyperrectangle (henceforth, called a non-basis polytope).   In either case the numerical problem amounts to drawing points uniformly from a non-basis polytope.

We have identified three methods that can be used to draw points from a non-basis polytope. We, however, find that only the third method is reliable.
\begin{enumerate}
\item Hit-and-Run (HR) sampling.  HR sampling uses Monte Carlo Markov Chain methods to draw points uniformly from the non-basis polytope $\Theta(h_{\text{rate}}^{\text{counter}}) \subset \mathbb{R}^d$.  The method is, however, numerically unstable if the non-basis polytope is thin. The contracted parameter space in the EAM algorithm converges to a polytope in $\mathbb{R}^{d-1}$ as the contraction counter increases  Therefore, HR sampling is unreliable for our problem.

\item Weighted average of vertices. In this method, the vertices of the contracted parameter space $\Theta(h_{\text{rate}}^{\text{counter}})$ are computed.  A randomly generated point can be generated from a random weighted average of the vertices.  Uniform weights do not guarantee that the point is uniformly drawn from $\Theta(h_{\text{rate}}^{\text{counter}})$.  This, never-the-less, does not violate convergence assumptions for the EAM algorithm provided that there is positive mass at all points $\theta \in \Theta(h_{\text{rate}}^{\text{counter}})$.  The algorithm that computes the vertices suffers from numerical issues as the parameter space becomes thin, and so this method is not appropriate for the CPI algorithm.

\item Draw-and-Discard sampling (DD).  The algorithm first draws points uniformly from a box $B \supset \Theta(h_{\text{rate}}^{\text{counter}})$.  It then discards any points that are not in $\Theta(h_{\text{rate}}^{\text{counter}})$.  The volume of $B$ relative to $\Theta(h_{\text{rate}}^{\text{counter}})$ must be small for this method to work well.  If not, then a large number of initial points are required in order to achieve a target number of points.  Therefore, the box $B$ needs to be carefully defined.
\end{enumerate}

In the current version of the CPI algorithm, the DD method only works for when $p$ is a basis vector and the parameter space is a non-basis polytope.  Modifications to the user-written function  \code{bound\_transform.m} are required.  We explain the modifications with an example.  The parameter space for DGP $8$ is the polytope:
\[
\Theta = \{\theta \in \mathbb{R}^5 : \theta_1 \in [0,1], \theta_2 \in [0,1],  \theta_k \in [0,\min\{\theta_1\,\theta_2\}], k=3,4,5\}.
\]
First, to run DD sampling set \code{KMSoptions.HR=0} (to use hit-and-run sampling set \code{KMSoptions\-.HR=1}).
To draw points from this space we use the draw-and-discard sampling method.  The file \code{bound\_transform.m} defines the box $B$ above.  It is not advised to set $B$ to be the parameter bounds $\theta_{LB}$ and $\theta_{UB}$, as the volume of this box relative to the contracted parameter space $\Theta(h_{\text{rate}}^{\text{counter}})$ quickly diverges.  The inputs of \code{bound\_transform} are: \code{LB\_in}, \code{UB\_in}, and \code{KMSoptions}.  The inputs \code{LB\_in} and \code{UB\_in} define the contracted parameter space (contracted in direction $p$). The outputs are the modified bounds \code{LB\_out} and \code{UB\_out} .  Points drawn from  $\{\theta \in \mathbb{R}^5 : LB_{\text{in}} \leq \theta \leq UB_{\text{in}}\}$ are unlikely to satisfy the polytope constraints.  In particular, if
\begin{align*}
 LB_{\text{in}} =
 \begin{bmatrix}
 0 \\
 0 \\
 0 \\
 0 \\
 0
 \end{bmatrix},
  \quad
   UB_{\text{in}} =
 \begin{bmatrix}
 10^{-4} \\
 1 \\
 1 \\
 1 \\
 1
 \end{bmatrix}
\end{align*}
then it is likely that components $3-5$ violate the condition $\theta_k \in [0,\min\{\theta_1\,\theta_2\}]$.  To resolve this issue the upper bound is modified, so that  $UB_{\text{out},1} = UB_{\text{in},1}$, $UB_{\text{out},2} = UB_{\text{in},2}$, and $UB_{\text{out},k} = \min\{UB_{\text{in},1},UB_{\text{in},2},UB_{\text{in},k}\}$ for $k=3,4,5$ (see \code{Lines 39-44} in  \code{bound\_transform.m}).  The lower bound is unchanged. The box $B$ defined by $LB_{\text{out}}$ and $UB_{\text{out}}$ contains the contracted parameter space and retains a good volume ratio.
The modifications to  \code{bound\_trans\-form.m} are model specific, and depend on the constraints $A_{\theta} \theta \leq b_{\theta}$.

If the parameter space is a polytope, then additional constraints for the linear program that computes $\hat c(\cdot)$ are required.  These constraints are determined by the user-specified function \code{Rho\_Polytope\_Box}.  Recall that we require $\lambda \in \sqrt{n}(\Theta - \theta)\cap \rho B^d$.  The constraint $\lambda_k \in [-\rho,\rho]$ is already included in \code{KMS\_33\_Coverage}.  For DGP 8, the following constraints need to be added:
\begin{align*}
&\lambda_k \leq \sqrt{n}(1 - \theta_k), k =1,2\\
&-\lambda_k \leq \sqrt{n}(0 - \theta_k), k = 1,2,3,4,5 \\
& -\lambda_1 + \lambda_k \leq - \sqrt{n}(- \theta_1 + \theta_k), k =3,4,5\\
& -\lambda_2 + \lambda_k \leq - \sqrt{n}(- \theta_2 + \theta_k), k =3,4,5.
\end{align*}
Observe that the constraint $-\lambda_1 + \lambda_k \leq - \sqrt{n}(- \theta_1 + \theta_k)$ is implied by the condition $\theta_k \leq \min\{\theta_1,\theta_2\}$.   These $S=13$ constraints are specified in \code{Rho\_Polytope\_Box}.  In the \code{CVXGEN} \code{C} code generator, we set $S=13$ for this DGP.
\end{appendices}

\end{document}